\documentclass[conference]{IEEEtran}

\usepackage{cite}
\usepackage{graphicx,color}
\ifCLASSINFOpdf

\else

\fi
\usepackage[cmex10]{amsmath}

\usepackage{bm}
\usepackage{amssymb}
\usepackage{algorithm}
\usepackage{algorithmic}
\usepackage{epstopdf}
\IEEEoverridecommandlockouts

\newtheorem{Def}{Definition}

\begin{document}
\title{Energy-Efficient Data Transmission with A Non-FIFO Packet}

\author{\IEEEauthorblockN{Qing Zhou, Nan Liu}
\IEEEauthorblockA{National Mobile Communications Research Laboratory, Southeast University, Nanjing 210096, China\\
Email:\{qingzhou and nanliu\}@seu.edu.cn}
%\thanks

}

\maketitle
%\linespread{1.6}

\begin{abstract}
This paper investigates the problem of energy-efficient packet transmission with a non-FIFO Packet over a point-to-point additive white Gaussian noise (AWGN) time-invariant channel under the feasibility constraints. More specifically, we consider the scenario where there is a packet that has a deadline that is earlier than that of the previously arrived packet.
% the individual deadlines of the packet sequence do not follow the order of their arrival instants
For this problem, the First-In-First-Out (FIFO) transmission mode adopted in the existing literatures is no longer optimal.
We first propose a novel packet split and reorder process which convert the inconsistency in the order of deadlines and arrival instants of the packet sequence into a consistent one. After the split and reorder process, the original problem considered in this paper is transformed into the problem of finding the optimal split factor.
%how to optimally split packet and how to optimally transmit packet under optimal split.
%Assuming a prior knowledge of packet size, arrival instants, individual deadlines as well as channel state information (CSI), we model the feasibility constraints data flow as cumulative curves.
We propose an algorithm that finds the split factor which consists of checking four possibilities by applying the existing optimal transmission strategy \emph{``String Tautening''} for FIFO packets. In addition, we prove the optimality of the proposed algorithm in the presence of a non-FIFO packet by exploiting the optimality properties of the most energy efficient transmission strategy. Based on the proposed optimal offline scheme, an efficient online policy which assumes causal arrival information is also studied and shown to achieve a comparable performance to the proposed optimal offline scheme.
\end{abstract}

\IEEEpeerreviewmaketitle
%\newpage
%\baselineskip=8.8mm

\section{Introduction}
Energy efficiency (EE) is an emerging issue for designing new communication systems to achieve significant energy savings, which will cut the operational costs as well as the  emission of carbon dioxide. References \cite{Berry2002},\cite{Uysal-Biyikoglu2002} showed that transmitting data flow with a low constant rate is an efficient method to reduce energy expenditure due to the fact that the transmission power is an increasing and strictly convex function of the transmission rate.
%On the other hand, since most of the current applications as well as services have been converting from traditional voice services into data services.
However, most of the current data services such as Voice over Internet Phone (VoIP) and video conferencing are often time-critical and delay-sensitive, therefore the Quality of Service (QoS) is an important factor which should be considered when we designing energy efficient realtime communication systems.

To this end, there have been many strategies put forth to address the energy-efficient transmission problems  \cite{Uysal-Biyikoglu2002,Zafer2009,Chen2008}. In \cite{Uysal-Biyikoglu2002}, the authors considered a transmission energy minimization problem for packet transmission with a single deadline constraint over a point-to-point additive white Gaussian noise (AWGN) time-invariant channel. A ``lazy scheduler'' was proposed as the optimal transmission strategy to achieve energy efficient packet transmission under the causality and deadline constraints. Generalizing \cite{Uysal-Biyikoglu2002} with respect to deadline constraints, \cite{Zafer2009,Chen2008} studied similar problems under individual deadline constraints: \cite{Zafer2009} posed the problem as a continuous time optimization and proposed a calculus approach to obtain the ``optimal departure curve'', which had a simple and appealing graphical visualization, which was named ``string tautening'' in \cite{Nan2014a}; in \cite{Chen2008}, a recursive optimal scheduling algorithm was put forward to find out the optimal policy to realize minimal energy consumption. In addition, \cite{Nan2014a,Jin2014} took the circuit power consumption into consideration and investigated energy efficient transmission of bursty data packet with individual deadlines under non-ideal circuit power consumption. In another relevant research field of energy harvesting, \cite{Xu2014,Bai2011,Ozel2011,Yang2012} studied the throughput maximizing problem or transmission time minimization problem for packet transmission subject to the causality constraint of energy arrivals and packet arrivals as well as the capacity constraint of the battery.

As far as we know, all previous work assumed that the packets are \emph{FIFO packets}, i.e.,  the individual deadlines of the data flow were consistent with the order of their arrival instants. However, in practical wireless communication systems, different applications and services have different requirements for packet delay, e.g., real-time voice or video and real-time games have high requirements on packet delay; while, buffered video streaming and TCP based services, such as www, ftp and e-mail, are less strict in terms of delay.
%Meanwhile, according to the different delay requirements of these different services, 3GPP TS 23.203\cite{3GPP2015} divided the Packet Delay Budget (PDB) into seven grades from UE to the Policy and Charging Enforcement Function (PCEF) in LTE, which are 50ms, 65ms, 75ms, 100ms, 150ms, 200ms and 300ms, respectively.
Therefore, it is very possible that a packet that has arrived later must depart before a packet that had arrived earlier. In other words, the consistency of the order of the deadlines and the arrival instants does not always hold.
For example, the heartbeat signals of typical Over The Top (OTT) application services such as WeChat and QQ, are likely to destroy the packet sequence consistency of deadlines and the arrival instants, since heartbeat signals tend to have higher latency requirements, i.e.,  the heartbeat packet's deadline is earlier than that of the previously arrived packet.
%Here, we refer to the packet that has a deadline earlier than that of the previously arrived packet as a \emph{non-FIFO packet}.
In addition, because the transmission cycle of the heartbeat packet is one minute or a few minutes, heartbeat packets are sufficiently apart from each other in time so we may study the scenario of one heartbeat packet inserted into a sequence of FIFO packets, which is the scenario considered in this paper.
% from the point of view of energy efficient transmission.
%Henceforth, this practical data flow application can be modeled as a sequence of FIFO packets inserting a \emph{non-FIFO packet}.
%However, all the energy efficient transmission strategies adopting FIFO mode as the transmission rule in \cite{Uysal-Biyikoglu2002,Zafer2009,Chen2008,Nan2014a,Jin2014} may not be applicable to the practical data flow application containing \emph{non-FIFO packet}. Yet, there is few work proposing optimal transmission strategy according to this situation.

More specifically, we consider a packet sequence transmission process over a point-to-point AWGN time-invariant channel and aim to minimize the total energy consumption of transmission under the feasibility constraints. We assume that the transmitter has a prior knowledge of the sizes, arrival instants and individual deadlines of the packets as well as the channel state information. We denote the
$i$-th packet that arrives as Packet $i$. In particular, we study the case where  there exists a Packet $j$ that has a deadline earlier than that of Packet $j-1$. This packet is called a \emph{non-FIFO packet}. The deadline and arrival instants of the other packets are in a consistent order.
%All the other packets are a \emph{non-FIFO packet} destroys the order consistency of the arrival instants and individual deadlines.
We first propose a novel packet split and reorder process  to restore the order consistency of the arrival instants and individual deadlines. This transforms the original energy minimization problem into the problem of finding the optimal split factor.
%Assuming a prior knowledge of packet size, arrival instants, individual deadlines as well as channel state information, we model the feasibility constraints of data flows as cumulative curves.
Then, we propose an algorithm that finds the split factor which consists of checking four possibilities by applying the existing optimal transmission strategy \emph{``String Tautening''} for FIFO packets. Based on the algorithm, we present the appealing string visualization of the proposed transmission strategy in the case of a non-FIFO packet, which can provide insight for designing online schedulers as well as further research on relevant problems. Next, the fact that the proposed algorithm can find the optimal split factor is proved by exploiting the optimality properties of the proposed strategy. Finally, a practical energy efficient online scheduler is studied.

\section{system model}
\subsection{Data Flow Model}
\begin{figure}
% Requires
\centering
\includegraphics[width=.4\textwidth]{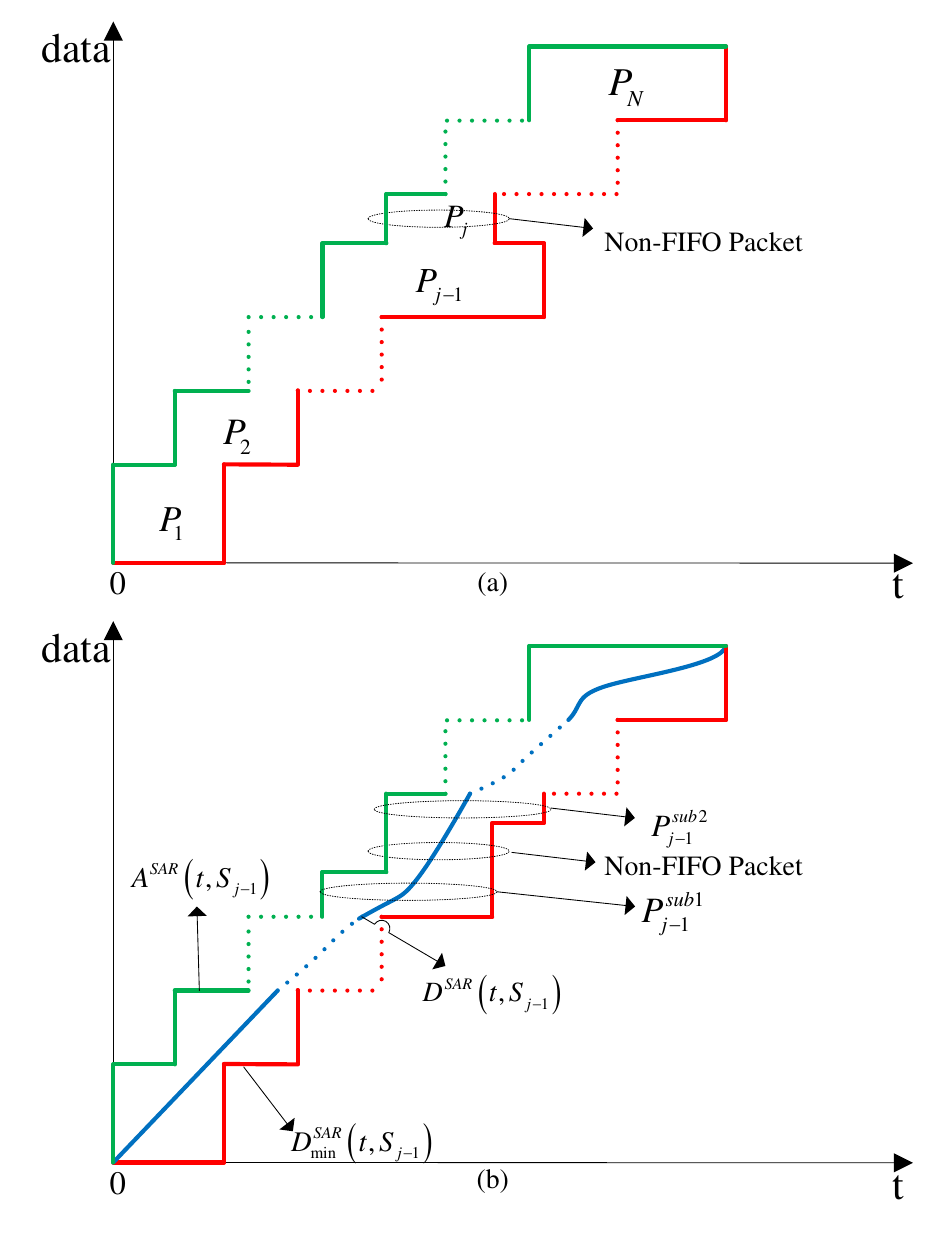}   \caption{(a). The arrival process of the original sequence $\mathcal{P}$; (b). Cumulative curves model for data flow of $\mathcal{P}^{SAR}(S_{j-1})$.}
\label{SystemModel}
\end{figure}

In this paper, we consider a point-to-point wireless link over an AWGN channel which is assumed to be time-invariant. As shown in fig.\ref{SystemModel}(a), there are $N$ packets randomly arriving at the transmitter buffer in sequence, and the set of the packets is denoted as $\mathcal{P} = \left\{ {{P_1},{P_2}, \ldots ,{P_N}} \right\}$. The key attributes of each packet can be expressed as ${I_i} = \left( {{B_i},{t_{a,i}},{t_{d,i}}} \right)$, $1 \le i \le N$, where ${B_i}$ is the size of the $i$-th packet, and ${t_{a,i}}$ and ${t_{d,i}}\left( { > {t_{a,i}}} \right)$ represent the corresponding arrival instant and the deadline of packet $P_i$, respectively.
%Henceforth, we can denote by $\mathcal{I}=\left\{I_1,\cdots,I_N\right\}$ as the key attributes set of the $N$ packets sequence.
For the offline transmission scheme, we assume that the key attributes of each packet as well as the channel state information (CSI) are a priori known at the transmitter, which is the assumption also made in \cite{Uysal-Biyikoglu2002,Zafer2009,Chen2008,Nan2014a}. For the online transmission scheme, we assume that the key attributes of each packet is known causally.
%
%
%In addition, we note that the CSI obtained by the transmitter is a constant since we just consider a time-invariant channel, hence, it only be used for the calculation of energy consumption and has no effect on the development of the optimal policy in this paper.

Without loss of generality, the first packet is assumed to arrive at instant 0, and the packets arrives in sequence, i.e., $0 = {t_{a,1}} < {t_{a,2}} <  \cdots  < {t_{a,N}}$. Previous works  \cite{Uysal-Biyikoglu2002,Zafer2009,Chen2008,Nan2014a,Jin2014} assumed that the deadlines of the packets follow the same order as the arrival times in the sense that ${t_{d,1}} < t_{d,2}<  \cdots  < t_{d,N}$. In this work, we consider the scenario where there exists a packet
${P_j}$, for some $j \in \left\{ 2,3, \cdots, N \right\}$, that has a deadline that is earlier than that of the previous packet $P_{j-1}$, while the deadlines of all the other packets are still in order, i.e., ${t_{d,1}} <  \cdots< t_{d,j-2}  < {t_{d,j}} < {t_{d,j - 1}} < t_{d,j+1}  \cdots  < T$, where $T$ is defined as $T = \max \left\{ {{t_{d,i}}|1 \le i \le N} \right\}$. We call packet $P_j$ the {\it{non-FIFO packet}}.

\subsection{Transmission Model}
We let $p\left( t \right)$ signify the transmission power at time $t$ when the transmission rate is $r\left( t \right)$. The relationship between $p\left( t \right)$ and $r\left( t \right)$ can be described using the function $f$ as:
\begin{equation}\label{PowerRate}
  p\left( t \right) = f\left( {r\left( t \right)} \right)
\end{equation}
where $f\left( \cdot \right)$ is a convex and increasing function defined on $[0, \infty]$. In addition, $p\left( x \right) \geq 0$ for all $t \in [0,\infty]$.

Shannon's capacity formula over an AWGN channel provides a typical example for the function $f$ as follows:
\begin{equation}\label{Shannon}
  r\left( t \right) = \frac{1}{2}\log \left( {1 + \frac{{p\left( t \right)}}{\sigma^2}} \right)
\end{equation}
where $\sigma^2$ is the variance of the channel noise. We may rewrite equation (\ref{Shannon}) as $p\left( t \right) = \sigma^2\left( {{2^{2r\left( t \right)}} - 1} \right)$. It can be easily verified that the expended power is  a convex and increasing function of the transmission rate. More examples of the function $f$ are provided in \cite{Uysal-Biyikoglu2002}.

\subsection{Problem Formulation}
The {\it{feasibility constraints}} of transmission include the causality constraint and the delay constraint: the causality constraint means that the transmitter cannot transmit the packets which have not yet arrived, and the delay constraint specifies that the packets must be completely transmitted before their individual deadlines. Henceforth, the objective of the paper is to find the transmission strategy that results in the minimum energy consumption under the {\it{feasibility constraints}}, i.e.,
\begin{align}
\min_{r(t)} \quad &E\left( {r\left( t \right):t \in \left[ {0,T} \right]} \right) = \int_0^T {f\left( {r\left( t \right)} \right)} dt\label{NanProb1}\\
 &\text{s. t.  }~~ \text{{\it{feasibility constraints}}}
\label{NanProb}
\end{align}

Throughout the paper, we use $x^-$ to denote $\lim_{\epsilon \rightarrow 0^+} (x-\epsilon)$.

\section{The Proposed Offline Strategy}

Before we proceed, we define a packet split and reorder process.
\begin{Def} \label{split_reorder}
Define a \emph{split and reorder} process as follows: given ${P_{_i}}, i\in \{1,2, \cdots,N\}$ with the key attributes $I_i= \left( {{B_i},{t_{a,i}},{t_{d,i}}} \right)$, assume package $P_j$ is the non-FIFO packet. For a \emph{split factor} $0 \le {S_{j-1}} \le {B_{j-1}}$, split ${P_{_{j-1}}}$ into two sub-packets: $P_{j-1}^{sub1}$ and $P_{j-1}^{sub2}$ with the sizes of ${S_{j-1}}$ and ${B_{j-1}} - {S_{j-1}}$, respectively, for some $S_{j-1} \in [0, B_{j-1}]$. The key attributes of package $P_{j-1}^{sub1}$ is $I_{j - 1}^{sub1} = \left( {{S_{j - 1}},{t_{a,j - 1}},{t_{d,j}}} \right)$, i.e., $P_{j-1}^{sub1}$ is a packet with size of $S_{j-1}$, and its arrival instant and deadline are the same as that of $P_{j-1}$ and $P_j$, respectively. The key attributes of package $P_{j-1}^{sub2}$ is  $I_{j - 1}^{sub2} = \left( {{B_{j - 1}} - {S_{j - 1}},{t_{a,j}},{t_{d,j - 1}}} \right)$, that is, $P_{j-1}^{sub2}$ has size $B_{j-1}-S_{j-1}$, and its arrival instant and deadline are the same as that of $P_j$ and $P_{j-1}$, respectively.
Now, consider the transmission of $N+1$ packets, reordered as follows: $\mathcal{P}^{SAR}(S_{j-1}) \triangleq \{P_1, \cdots, P_{j-1}^{sub1}, P_j, P_{j-1}^{sub2}, \cdots, P_N\}$, where the superscript $SAR$ is short for ``split and reorder''. The arrival curve, departure curve and minimum departure curve of $\mathcal{P}^{SAR}(S_{j-1})$ are denoted as $A^{SAR}\left( t,{S_{j - 1}} \right)$, $D^{SAR}(t, S_{j-1})$ and $D^{SAR}_{\min }\left( t,S_{j - 1} \right)$, respectively.
\end{Def}

Similar to \cite{Zafer2009}, to intuitively describe the data flow model after \emph{split and reorder} process in this system, i.e., $\mathcal{P}^{SAR}(S_{j-1})$, we adopt a cumulative curves methodology. The model consists of an arrival curve, a departure curve and a minimum departure curve, which are defined as follows:

\begin{Def}
The {\it{arrival curve}} $A^{SAR}\left( {t},S_{j-1} \right), t \in \left[ {0,T} \right]$ is defined as the total number of bits that have arrived in $\left[ {0,t} \right]$.
\end{Def}

\begin{Def}
The {\it{departure curve}} $D^{SAR}\left( {t},S_{j-1} \right), t \in \left[ {0,T} \right]$ is defined as the total number of bits that have departed in $\left[ {0,t} \right]$.
\end{Def}

\begin{Def}
The {\it{minimum departure curve}}  ${D^{SAR}_{\min }}\left( t,S_{j-1} \right), t \in \left[ {0,T} \right]$ is defined as the minimum number of bits that must depart by $t$ to meet the deadline constraints of the packets.
\end{Def}

Based on Definitions 2-4, a {\it{departure curve}} $D^{SAR}\left( {t},S_{j-1} \right)$, $t \in \left[ {0,T} \right]$, $S_{j-1} \in \left[ {0,B_{j-1}} \right]$, is feasible if it satisfies the feasibility constraints, which can be described as:
\begin{equation}\label{FeasibleCondition}
  {D^{SAR}_{\min }}\left( t ,S_{j-1}\right) \le D^{SAR}\left( t,S_{j-1} \right) \le A^{SAR}\left( t ,S_{j-1}\right)
\end{equation}

Fig. \ref{SystemModel}(b) depicts the cumulative curves model of data flow after \emph{split and reorder} process in this paper, where the dashed curves denote the cumulative curves of other omitted packets. The minimum departure curve is below the arrival curve. Since the departure curve needs to be feasible, it must satisfy (\ref{FeasibleCondition}), and consequently, it lies in between the arrival curve and the minimum departure curve.
%The dashed curves denote the cumulative curves of other omitted packets.

Note that the rate of any departure curve at time $t$ is equal to its derivative, i.e., $r( t ) = {D^{SAR}}'( t,S_{j-1}) $. Hence, the energy minimization problem for the sequence after the \emph{split and reorder} process, i.e., $\mathcal{P}^{SAR}(S_{j-1})$, can be formulated as follows:
\begin{align}
&\min_{S_{j-1}} \min_{D^{SAR}(t,S_{j-1})} E\left( D^{SAR}\left( t,{S_{j - 1}} \right) \right) =\nonumber\\
&\min \int_0^T f\left( {{D^{SAR}}'\left( {t,{S_{j - 1}}} \right)} \right) dt \label{equalformulation} \\
&\text{s. t. } (\ref{FeasibleCondition}),  ~ 0 \le S_{j - 1} \le B_{j - 1},t \in \left[ 0,T \right]\label{equalformulation1}
\end{align}

Note that the sequence of packets $\mathcal{P}^{SAR}(S_{j-1})$ has no non-FIFO packet, i.e., the arrival instants and the deadlines of the packets follow the same order. The optimal strategy when there is {\it{no}} non-FIFO packet has been found in \cite{Zafer2009}, which is called the ``String Tautening'' scheduler \cite{Nan2014a}.

\subsection{The Optimal Strategy Without Non-FIFO Packets}

In this section, we review the existing results of packet sequence transmission without non-FIFO packets. In this case, the individual deadlines of the packets are in the same order as their arrival times. Henceforth, packets wait in queue and are transmitted following the FIFO rule. The corresponding optimal strategy  \cite{Zafer2009} is presented in Algorithm \ref{String Tautening}:
\begin{algorithm}
\caption{\itshape ``String Tautening'' for FIFO packets}
\begin{algorithmic}[1]\label{String Tautening}
\STATE Setting ${t_0} = 0$, $D_{opt}(0)=0$ and $D_{opt}(T)=D_{min}(T)$, beginning with the starting point $({t_0} = 0, D_{opt}(0)=0)$ to obtain $D_{opt}(t)$ in a recursive way;
\STATE Finding out ${\beta _0} = \inf  {F_A} = \sup{F_{{D_{\min }}}}$ and obtaining the optimal line segment ${L_0}$, where ${F_A}\left( {{F_{{D_{\min }}}}} \right)$ represents the positive slope set of radials which starts at the starting point and intersect $A\left( t \right)\left( {{D_{\min }}\left( t \right)} \right)$ first.
\STATE Obtaining the first intersection instant ${t_1}$ that ${L_0}\left( {{t_1}} \right) = {A}\left( {{t_1}} \right)$ or ${L_0}\left( {{t_1}} \right) = {D_{\min }}\left( {{t_1}} \right)$ and setting ${D_{opt}}\left( t \right) = {L_0}\left( t \right)$, where $t \in \left( {{t_0},{t_1}} \right]$;
\STATE If ${t_1} = T$, then the algorithm terminates; else, replace starting point by the new starting point $\left( {{t_1},{D_{opt}}\left( {{t_1}} \right)} \right)$ and repeat steps 2 and 3.
\end{algorithmic}
\end{algorithm}

Note that $A(t)$, $D(t)$ and $D_{min}(t)$ in algorithm \ref{String Tautening} are the arrival curve, departure curve and minimum departure curve, respectively. In addition, reference \cite{Zafer2009} provided a \textit{string visualization} of the obtained optimal strategy using Algorithm \ref{String Tautening}:  we tie one end of a string at the starting point $\left( {0,0} \right)$ and pass the other end through the ending point $\left( {T,{D_{\min }}\left( T \right)} \right)$. If we pull the string tightly between $A\left( t \right)$ and ${D_{\min }}\left( t \right)$, then the trajectory of the string is the optimal departure curve that results in the least amount of transmission energy expenditure.
%${D_{opt}}\left( t \right),t \in \left[ {0,T} \right]$.

Note that for a given ${S_{j - 1}}$, we can find the optimal departure curve, denoted as $D_{opt}^{SAR}(t, S_{j - 1})$, for $\mathcal{P}^{SAR}(S_{j-1})$ according to Algorithm 1  since there are no non-FIFO packets in $\mathcal{P}^{SAR}(S_{j-1})$.  The optimal departure curve found for $\mathcal{P}^{SAR}(S_{j-1})$, i.e., $D_{opt}^{SAR}(t, S_{j - 1})$, has the following property: if there exist any two points of $D^{SAR}(t, S_{j - 1})$ that can be joined by a straight line segment, then replacing the relevant portion of $D^{SAR}(t, S_{j - 1})$ by the straight line segment can reduce energy consumption. Henceforth, whenever allowable, it is optimal to transmit the packets at a constant rate \cite[Theorem 1]{Zafer2009}.  We call this the \emph{constant rate property}. The constant rate policy implies that for the case of FIFO packets, each packet should be transmitted with a constant rate \cite{Uysal-Biyikoglu2002,Zafer2009}.

\subsection{The Proposed Strategy with A Non-FIFO Packet}
 \label{proposed_strategy}
The proposed strategy when there is a non-FIFO packet consists of checking 4 possibilities and applying Algorithm 1 under each possibility. We describe the procedure below and the optimality of the procedure will be proved in the next section.

\subsubsection{Possibility 1}
%First, for the \emph{non-FIFO packet} ${P_j},j \in \left\{2, 3,\cdots, N\right\}$, that satisfies ${t_{d,j}} < {t_{d,j - 1}}$, we
Let $S_{j-1}=0$, and perform the split and reorder process, i.e., Packet $j-1$ is not split, and furthermore, it is transmitted after Packet $j$. The corresponding packets considered after the split and reorder process is $\mathcal{P}^{SAR}(0)=\{P_1, \cdots, P_j, P_{j-1}, P_{j+1}, \cdots, P_N\}$, where the key attributes of $P_{j-1}$ is changed to $(B_{j-1}, t_{a,j}, t_{d,j-1})$ while the key attributes of all the other packets remain the same.
%Note that the deadline of the packets are in the same order as the arrival instants for $\mathcal{P}^{new}(0)=\{P_1, \cdots, P_j, P_{j-1}, P_{j+1}, \cdots, P_N\}$.
%
Use Algorithm \ref{String Tautening} to find the optimal departure curve $D^{SAR}_{opt}(t,0)$ for $\mathcal{P}^{SAR}(0)$.

Define the following condition:
\begin{align}
{D^{SAR}_{opt}}'(b^{-}_{j-2},0)\ge{D^{SAR}_{opt}}'(b^{-}_{j-1},0) \label{condition02}
\end{align}
where $b_{j-2}$ and $b_{j-1}$ are the time instants when Packets $P_{j-2}$ and $P_{j-1}$ are completely transmitted, respectively, i.e., $D^{SAR}_{opt}(b_{j-2},0)=\sum_{i=1}^{j-2}B_i$ and $D^{SAR}_{opt}(b_{j-1},0)=\sum_{i=1}^{j}B_i$. Henceforth, ${D^{SAR}_{opt}}'(b^{-}_{j-2},0)$ and ${D^{SAR}_{opt}}'(b^{-}_{j-1},0)$ are the transmission rate of Packets $P_{j-2}$ and $P_{j-1}$, respectively. Condition (\ref{condition02}) means that the transmission rate of the previous packet of $P_j$, i.e., $P_{j-2}$, is no smaller than that of the next packet of $P_j$, i.e., $P_{j-1}$.

If the condition in (\ref{condition02}) is satisfied, we adopt the transmission strategy of  $D^{SAR}_{opt}(t,0)$ as the solution to the problem (\ref{NanProb1}), otherwise, we continue to check Possibility 2.

% Furthermore, $D^M_{opt}(t)$ is the optimal strategy as we will prove in Section \ref{nan_optimal}.
\subsubsection{Possibility 2}
Let $S_{j-1}=B_{j-1}$, and perform the packet split and reorder process, i.e., the packet $P_{j-1}$ is not split and furthermore, it is fully transmitted before the transmission of packet $P_j$. The corresponding packets considered after the split and reorder process is $\mathcal{P}^{SAR}(B_{j-1})=\{P_1, \cdots, P_{j-1}, P_j, P_{j+1}, \cdots, P_N\}$, where the key attributes of the packet $P_{j-1}$ is changed to $(B_{j-1}, t_{a,j-1}, t_{d,j})$ while the key attributes of all the other packets remain the same.
Use Algorithm \ref{String Tautening} to find the optimal departure curve $D^{SAR}_{opt}(t,B_{j-1})$ for $\mathcal{P}^{SAR}(B_{j-1})$.

Define the following condition:
\begin{align}
{D^{SAR}_{opt}}'(b^{-}_{j-1},B_{j-1}) \le {D^{SAR}_{opt}}'(b^{-}_{j+1},B_{j-1}) \label{condition03}
\end{align}
where $b_{j-1}$ and $b_{j+1}$ represent the instants that Packet $P_{j-1}$ and $P_{j+1}$ is completely transmitted, respectively, i.e., $D^{SAR}_{opt}(b_{j-1},B_{j-1})=\sum_{i=1}^{j-1}B_i$ and $D^{SAR}_{opt}(b_{j+1},B_{j-1})=\sum_{i=1}^{j+1}B_i$. Henceforth, ${D^{SAR}_{opt}}'(b^{-}_{j-1},B_{j-1})$ and ${D^{SAR}_{opt}}'(b^{-}_{j+1},B_{j-1})$ are the transmission rates of $P_{j-1}$ and $P_{j+1}$, respectively. Opposite to condition (\ref{condition02}) in Possibility 1, condition (\ref{condition03}) means that the transmission rate of the previous packet of $P_j$, i.e., $P_{j-1}$, is no larger than that of the next packet of $P_{j}$, i.e., $P_{j+1}$.

If the condition in (\ref{condition03}) is satisfied, we adopt the transmission strategy of  $D^{SAR}_{opt}(t,B_{j-1})$ as the solution to the problem (\ref{NanProb1}), otherwise, we go on and check Possibility 3.
\subsubsection{Possibility 3}
Merge ${P_j}$ and its previous packet ${P_{j - 1}}$ together to form a new packet $P_{j - 1}^M$ with the key attribute $I^M_{j-1}= \left( {{B_{j - 1}} + {B_j},{t_{a,j - 1}},{t_{d,j - 1}}} \right)$, i.e., the new packet $P_{j - 1}^M$ is of size $B_{j - 1} + {B_j}$, and the corresponding arrival instant and deadline are the same as that of $P_{j-1}$.
We denote the new packet sequence as $\mathcal{P}^M=\{P_1, P_2, \cdots, P_{j-2}, P_{j-1}^M, P_{j+1}, \cdots, P_N\}$. Note that there are no non-FIFO packets in $\mathcal{P}^M$, and the individual deadline of all the packets are in the same order as their arrival instants. Furthermore, denote the new arrival curve and minimum departure curve of the new packet sequence as $A^M\left( t \right)$ and ${D^M_{\min }}\left( t \right)$, respectively. Use Algorithm \ref{String Tautening} to find the optimal departure curve, denoted as $D_{opt}^M(t)$, for $\mathcal{P}^M$.

To transmit using $D^M_{opt}(t)$ for our original problem, we perform a packet split and reorder process with respect to $S_{j-1}=D_{opt}^M(t_{d,j})-\sum_{i=1}^{j-2} B_j-B_j$, which means that the packet $P_j$ has just finished transmitting by its deadline $t_{d,j}$ and the amount of data in packet $P_{j-1}$ that has been transmitted by time $t_{d,j}$ is equal to the amount of data transmitted by the strategy $D_{opt}^M$ at time $t_{d,j}$ minus the amount of data in packets $P_1, P_2, \cdots, P_{j-2}, P_j$. Check the following condition:
\begin{equation}\label{condition01}
  \left\{ {\begin{array}{*{20}{c}}
  D^M_{opt}\left( {{t_{a,j}}} \right)\le A^{SAR}\left( {{t_{a,j}},{S_{j - 1}}} \right)\\
  0<S_{j-1}<B_{j-1}\\
  \end{array}}
  \right.
\end{equation}
where the first component is the causality constraint, i.e., the amount of data transmitted using strategy $D^M_{opt}(t)$ by the time $t_{a,j}$ can not be more than the amount of data that has arrived for the split process $\mathcal{P}^{SAR}(S_{j-1})$.

If condition (\ref{condition01}) is true, it means that we can perform a packet split with respect to the split factor  $S_{j-1}$ and transmit using the strategy $D^M_{opt}(t)$, which will satisfy the feasibility constraint. On the other hand,
if the condition (\ref{condition01}) is not satisfied,
%it means that though the energy consumption of $D^M_{opt}(t)$ is minimal, it
it means that $D^M_{opt}(t)$ does not satisfy the feasibility constraint. Hence, if condition (\ref{condition01}) is true, we adopt the transmission strategy $D^M_{opt}(t)$ as the solution to the problem in (\ref{NanProb1}), otherwise, we have Possibility 4.

\subsubsection{Possibility 4}

If none of the above three possibilities are satisfied, we reserve the time from $t_{a,j}$ to $t_{d,j}$ exclusively  for the transmission of Packet $P_j$, i.e., as soon as the packet $P_j$ arrives, we start its transmission and it finishes transmission just in time by the deadline $t_{d,j}$. The optimal strategy for the remaining packets, i.e., $P_1, P_2, \cdots, P_{j-1}, P_{j+1}, \cdots, P_N$ can be found by Algorithm \ref{String Tautening} by removing the time period $[t_{a,j}, t_{d,j}]$ as this period has been reserved exclusively for packet $P_j$.

More specifically, consider the remaining $N-1$ packets as a new packet sequence denoted as $\mathcal{P}^R = \left\{ {{P_1}, \ldots ,{P_{j - 1}},{P_{j + 1}}, \ldots ,{P_N}} \right\}$, where the arrival instants and deadlines of the $N-1$ packets have changed to
\begin{equation}\label{equalarrivalinstant}
t_{a,i}^R = \left\{ {\begin{array}{*{20}{c}}
   {{t_{a,i}},} & {{t_{a,i}} \le {t_{a,j}};}  \\
   {{t_{a,j}},} & {{t_{a,j}} < {t_{a,i}} \le {t_{d,j}};}  \\
   {{t_{a,i}} - \left( {{t_{d,j}} - {t_{a,j}}} \right),} & {{t_{a,i}} > {t_{d,j}}.}  \\
\end{array}} \right.
\end{equation}
and
\begin{equation}\label{equaldeadline}
t_{d,i}^R = \left\{ {\begin{array}{*{20}{c}}
   {{t_{d,i}},} & {{t_{d,i}} \le {t_{a,j}};}  \\
   {{t_{a,j}},} & {{t_{a,j}} < {t_{d,i}} \le {t_{d,j}};}  \\
   {{t_{d,i}} - \left( {{t_{d,j}} - {t_{a,j}}} \right),} & {{t_{d,i}} > {t_{d,j}}.}  \\
\end{array}} \right.
\end{equation}
where we denote $t_{a,i}^R$ and $t_{d,i}^R$ as the respective arrival instants and deadline of the new packet sequence $\mathcal{P}^R$. To explain (\ref{equalarrivalinstant}) and (\ref{equaldeadline}) in words, the arrival or departure times of the packets that are before time $t_{a,j}$ remains the same, arrival and departure times that are between the arrival instant and deadline of packet $P_j$ is set to be the arrival instant of packet $P_j$, and the arrival and departure times that are after the departure instant of packet $P_j$ is shifted by the amount of $t_{d,j} - t_{a,j}$ to an earlier time.

Note that for the packet sequence $\mathcal{P}^R$, the individual deadline of all the packets are in the same order as their arrival instants. Henceforth, use Algorithm \ref{String Tautening} to obtain optimal departure curve $D^R_{opt}(t)$ for $\mathcal{P}^R$. As a result, the split factor is $S_{j-1}=D^R_{opt}(t_{a,j})-\sum_{i=1}^{j-2} B_i$, and the strategy we adopt as the solution to the problem in (\ref{NanProb1}) is $D^{P_4}(t)$, where
\begin{align}
&D^{P_4}(t)= \nonumber\\
&\left\{
\begin{array}{ll}
D^R_{opt}(t) & \text{ if } t \leq t_{a,j}\\
D^R_{opt}(t_{a,j})+\frac{B_j}{t_{d,j}-t_{a,j}}(t-t_{a,j}) & \text{ if } t_{a,j} < t< t_{d,j} \\
D^R_{opt}(t-(t_{d,j}-t_{a,j}))+B_j &\text{ if } t \geq t_{d,j}
\end{array}
\right.\
\end{align}

\subsection{String Visualization of the Proposed Strategy}

\begin{figure}[htb]
 %Requires
\centering
\includegraphics[width=.45\textwidth]{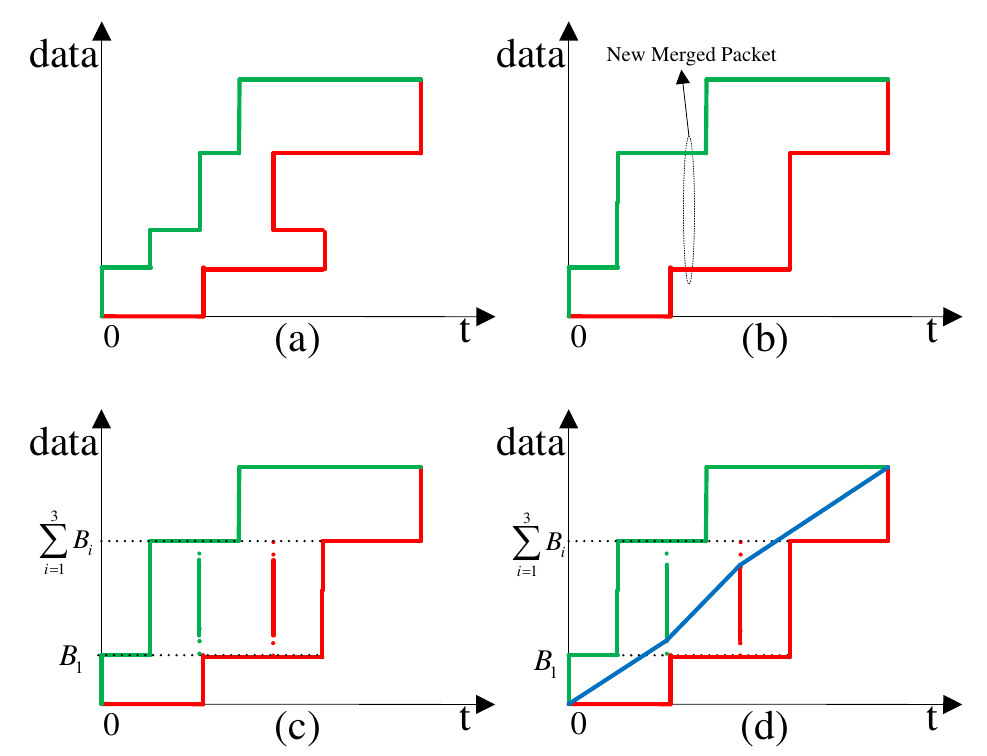}
\caption{Illustration of String Visualization of the Proposed Strategy of 4 packets.}
\label{Stringvisualization}
\end{figure}

Recall the packet sequence $\mathcal{P}^M$ defined in Possibility 3.
% First, plot the arrival curve and minimum departure curve of for the original packet sequence according to the order of the arrival instants, as shown in Fig \ref{Stringvisualization}.a.
First, plot the arrival curve and minimum departure curve of for the packet sequence  $\mathcal{P}^M$, as shown in Fig \ref{Stringvisualization}(b), where Figure \ref{Stringvisualization}(a) depict the arrival process for original sequence of packets $\mathcal{P}$ where there is a non-FIFO packet. Secondly, consider two horizontal walls located at the heights of  $\text{data}=\sum_{i=1}^{j-2} B_i$ and $\text{data}=\sum_{i=1}^j B_j$, respectively. The two horizontal walls are depicted by the black dotted line in  Fig \ref{Stringvisualization}(c).  Thirdly, add two vertical line segment of length $B_j$ in the positions of $t=t_{a,j}$ and $t=t_{d,j}$, respectively. The two line segment can slide vertically between the two horizontal walls, see Fig \ref{Stringvisualization}(c). The sliding of the two vertical lines between the two horizontal walls is due to the fact that $S_{j-1}$ can vary between $[0, B_{j-1}]$. Note that the slide is synchronous between the two line segments, i.e., during the slide, the second (vertical axis) coordinate of the down end points of the two line segments stay the same. The \emph{String Visualization} can be described as: tie one end of a string at the origin and  pass the other end between the two vertical line segments and then connect the point $( T,D_{min}^{M}(T))$. Next, pull the string as tightly as possible, pulling the string will cause the two vertical line segment to slide between the two horizontal walls. The trajectory of the tightest string indicate the optimal departure curve. Suppose that the second (vertical axis) coordinate of the down end points of the two vertical segments are $B$, then the optimal split factor is found to be $S_{j-1}^{opt}=B-\sum_{i=1}^{j-2} B_i$.

\section{The Optimality of the Proposed Strategy}
In this subsection, we prove the optimality of the proposed strategy in Section \ref{proposed_strategy} in the presence of a non-FIFO packet. We spit the proof into steps using several lemmas.

We first show that in the presence of a non-FIFO packet, the split and reorder process defined in Definition \ref{split_reorder} is an optimal strategy. This is stated more rigorously in the following lemma.

\upshape  \textbf{Lemma 1:}\label{split_optimal} {\itshape The energy minimization problem considered in (\ref{NanProb1}) and (\ref{NanProb}) is equivalent to the problem in (\ref{equalformulation}) and (\ref{equalformulation1}) .}

\upshape  \textbf{Proof:} Please See Appendix \ref{app_split_optimal}.\hfill $\Box$

For the optimization problem described in (\ref{equalformulation}),  ${S_{j - 1}}$ as well as $D^{SAR}\left( {t,{S_{j - 1}}} \right)$ are the two variables of this problem. For a given ${S_{j - 1}}$, we can find the optimal $D^{SAR}\left( {t,{S_{j - 1}}} \right)$, i.e.,  $D_{opt}^{SAR}(t, S_{j - 1})$, according Algorithm 1 since there are no non-FIFO packets in $\mathcal{P}^{^{SAR}}(S_{j-1})$. Hence, the problem of finding the optimal strategy in terms of minimum energy consumption reduces to the problem of finding the optimal split factor, denoted as $S^{opt}_{j-1}$.

We know that $S^{opt}_{j-1} \in [0, B_{j-1}]$, which means that $S^{opt}_{j-1}$ can take one of the following cases: $1) S^{opt}_{j-1}=0$; $2) S^{opt}_{j-1}=B_{j-1}$; $3) S^{opt}_{j-1} \in (0,B_{j-1})$.

Before we proceed, first recall the definition of $D^{SAR}_{opt}(t, S_{j - 1})$, which is the optimal departure curve found for $\mathcal{P}^{SAR}(S_{j-1})$ using Algorithm 1, where $\mathcal{P}^{SAR}(S_{j-1})$ is the resulting sequence of packets after the split and reorder process.

\subsubsection{Necessary and sufficient conditions for $S^{opt}_{j-1}=0$}
The necessary and sufficient conditions for $S^{opt}_{j-1}=0$ is stated in the following lemma:

\upshape  \textbf{Lemma 2:}{\itshape ${S^{opt}_{j - 1}} = 0$ if and only if the condition in (\ref{condition02}) is satisfied for $D^{SAR}_{opt}(t, 0)$.}\label{s0}

\upshape  \textbf{Proof:} Please See Appendix \ref{app_S=0/B}.\hfill $\Box$

From Lemmas 1 and 2, we see that to find the energy minimization problem of the given packets  $\mathcal{P}$, we can first use Algorithm 1 to obtain the optimal departure curve, i.e., $D^{SAR}_{opt}(t, 0)$, for $\mathcal{P}^{SAR}(0)$. If the condition stated in (\ref{condition02}) is satisfied, then $S^{opt}_{j-1}=0$. Otherwise, $S^{opt}_{j-1} \neq 0$.

\subsubsection{Necessary and Sufficient Conditions for $S^{opt}_{j-1}=B_{j-1}$}
The necessary and sufficient conditions for $S^{opt}_{j-1}=B_{j-1}$ is stated in the following lemma:

\upshape  \textbf{Lemma 3:}\label{s0} {\itshape ${S^{opt}_{j - 1}} = B_{j-1}$ if and only if the condition in (\ref{condition03}) is satisfied for $D^{SAR}_{opt}(t, B_{j-1})$.}

\upshape  \textbf{Proof:} Please See Appendix \ref{app_S=0/B}.\hfill $\Box$

To summarize the two cases above, we have shown that for a given packet sequence $\mathcal{P}$, we can first use Algorithm 1 to find the optimal departure curve $D^{SAR}_{opt}(t,0)$. If the condition in (\ref{condition02}) is satisfied, then $S^{opt}_{j-1}=0$. Otherwise, $S^{opt}_{j-1} \neq 0$ and we move on to the next possibility: use Algorithm 1 to find the optimal departure curve $D^{SAR}_{opt}(t,B_{j-1})$. If the condition in (\ref{condition03}) is satisfied, then $S^{opt}_{j-1}=B_{j-1}$, otherwise, $S^{opt}_{j-1}$ is not equal to $B_{j-1}$ either, which means that the optimal $S_{j-1}$ must be in $(0,B_{j-1})$.

\subsubsection{The Case where $S^{opt}_{j-1} \in (0,B_{j-1})$}

\begin{figure}[htpb]
% Requires
\centering
\includegraphics[width=.5\textwidth]{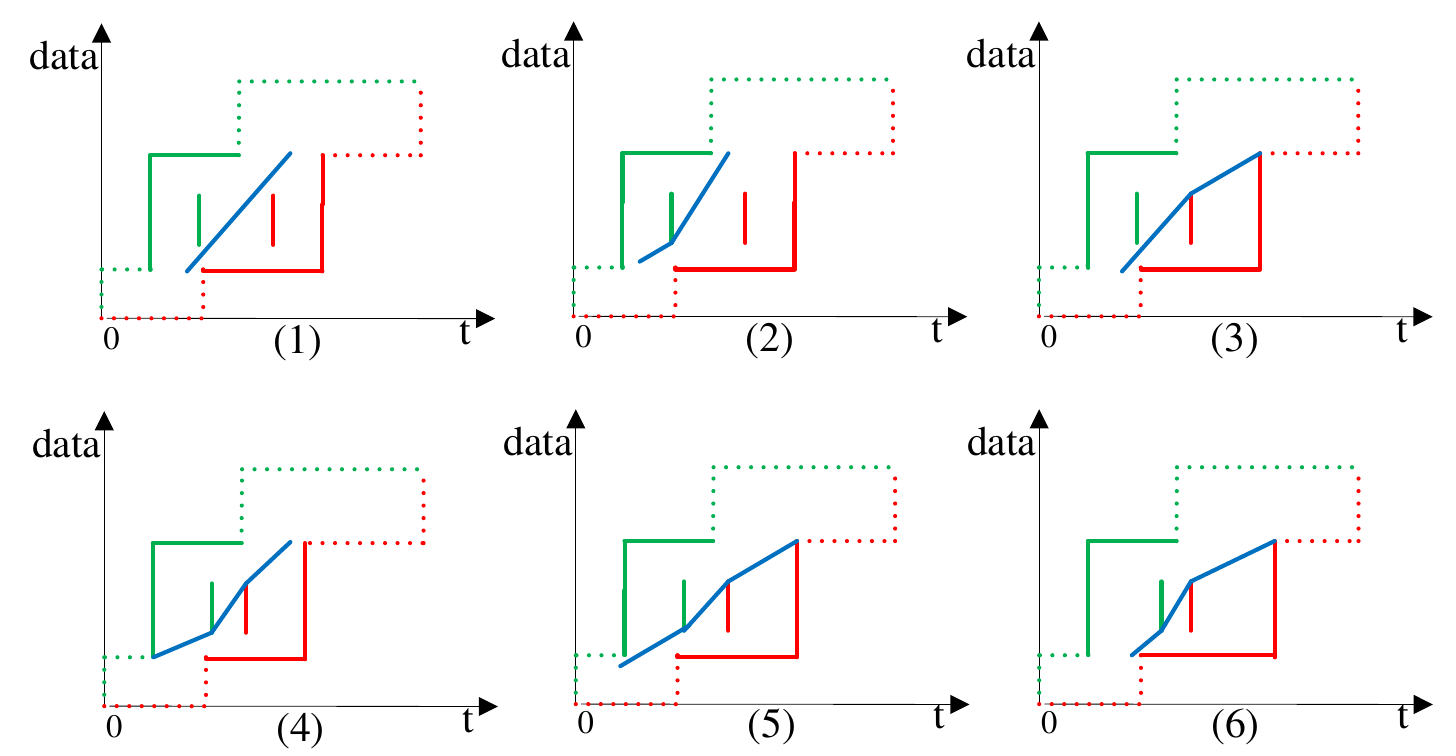}   \caption{All possible scenarios of $D^{SAR}_{opt}(t,S_{j-1})$ when $S_{j-1} \in (0,B_{j-1})$.}
\label{All possible cases}
\end{figure}

Due to the constant rate property, for the case of $S^{opt}_{j-1} \in (0,B_{j-1})$, the optimal departure curve can only take one of the six scenarios depicted in Figure \ref{All possible cases}.

Next, we show another optimality criterion for the optimal departure curve. Given that $S^{opt}_{j-1} \in (0,B_{j-1})$, which means that packet $j-1$ needs to be split into two sub-packets, the optimal departure curve must satisfy that the two sub-packets of $P_{j-1}$ are transmitted with equal rate. This is stated more rigorously in the following lemma.

\upshape  \textbf{Lemma 4:} {\itshape ${S^{opt}_{j - 1}} \in (0,B_{j-1})$ if and only if the following condition is satisfied:
\begin{equation}\label{equalrate}
  {D^{SAR}_{opt}}^\prime \left( {{t_{sub1}},{{S}^{opt}_{j - 1}}} \right) = {D^{SAR}_{opt}}^\prime \left( {{t_{sub2}},{{S}^{opt}_{j - 1}}} \right)
\end{equation}
$t_{sub1}\in({a^{sub1}_{j-1}},{b^{sub1}_{j-1}}]$ and $t_{sub2}\in({a^{sub2}_{j-1}},{b^{sub2}_{j-1}}]$, where $\left( {{a^{sub1}_{j-1}},{b^{sub1}_{j-1}}} \right]$ and $\left( {{a^{sub2}_{j-1}},{b^{sub2}_{j-1}}} \right]$ are the actual transmission duration of two sub-packets $P_{j - 1}^{sub1}$ and $P_{j - 1}^{sub2}$, respectively.}

\upshape  \textbf{Proof:} Please See Appendix \ref{app_0<S<B}.\hfill $\Box$

This means that among the six scenarios drawn in Figure \ref{All possible cases}, only Scenarios (1) and (5) can be the optimal departure curve. Scenario (1) corresponds to the case where the transmission rate of the $P_{j-1}^{sub1}$, $P_{j-1}^{sub2}$ and $P_j$ are all the same. This is the case where Packets $j-1$ and $j$ are transmitted with the same rate, thus, it corresponds to Possibility 3 in the proposed algorithm, where Packtes $j-1$ and $j$ are merged into one packet $P_{j - 1}^M$ with the key attribute $I^M_{j-1}= \left( {{B_{j - 1}} + {B_j},{t_{a,j - 1}},{t_{d,j - 1}}} \right)$. Find the optimal departure curve, i.e., $D_{opt}^M(t)$, for $\mathcal{P}^M=\{P_1, P_2, \cdots, P_{j-2}, P_{j-1}^M, P_{j+1}, \cdots, P_N\}$ using Algorithm 1. If $D_{opt}^M(t)$ satisfies (\ref{condition01}), it means that we have Scenario (1), and $D_{opt}^M(t)$ is obviously the optimal departure curve, as this is the optimal departure curve for the case of $t_{d,j}=t_{d,j-1}$, which is a less stringent condition than that of the original problem of $t_{d,j}<t_{d,j-1}$, and therefore, the minimum energy transmitted in the case of $t_{d,j}=t_{d,j-1}$ is a lower bound on the amount of energy transmitted for the problem considered in this paper.

If $D_{opt}^M(t)$ does not satisfy (\ref{condition01}), it means that Scenario (1) is not feasible, and we are left with the only possibility of Scenario (5) as the optimal departure curve, where the time between $t_{a,j}$ to the time $t_{d,j}$ is solely used for the transmission of Packet $j$, which corresponds to Possibility 4.

This concludes the proof of the optimality of the proposed strategy in Section  \ref{proposed_strategy}.

\section{Online Scheduler}
In Section III, we proposed the optimal offline algorithm assuming that the arrival information of all the packets during the duration $[0,T]$ was a priori known at the transmitter. This is an ideal assumption and the performance obtained serves as a lower bound on the total energy consumption in practical scenarios, where the arrival information of the packets is causal. In this section, we develop a heuristic online algorithm based on the proposed optimal offline scheme. The core concept of the online algorithm is to adopt the optimal offline algorithm to schedule all the packets have arrived so far, and reschedule when a new packet arrives. To be specific, when $t_{a,1}=0$, we adopt Algorithm 1 to obtain the optimal transmission strategy according to the information of the first arrival packet until the new packet arrives at $t_{a,2}$. Then we consider the new arrival instant $t_{a,2}$ as the starting point of the packet sequence, meanwhile, the {\it{Arrival Curve}} at $t_{a,2}$: $A^{ON}(t_{a,2})$, where ``ON'' is short for ``Online'', is the sum size of the remaining packets in the buffer and the new arrival packet. If the newly arrived packet is a FIFO packet, then, we schedule all the packets based on Algorithm 1 according to the deadlines and the packet size of the remaining packets in the buffer as well as the newly arrived packet; if the newly arrived packet is a Non-FIFO packet, similarly, we schedule all the packets based on the proposed optimal offline algorithm in section III according to the relevant deadlines and packet sizes. Note that the deadlines of the {\it{Minimum Departure Curve}} $D^{ON}_{min}(t)$ contains those of the remaining packets in the buffer and the newly arrived packet and does not contain the past deadlines. We go on scheduling packets until all the packets have arrived and are completely transmitted.

\section{Numerical Results}
We consider a time invariant AWGN channel with channel gain $|h|^2 = 2$  and the AWGN noise $\sigma^2 =1$ \cite{Nan2014a}. The bandwidth $W = 1$ KHz \cite{Nan2014a} and the FIFO packets are of size $B = 1$ KB . The energy-rate function is indicated by the Shannon capacity formula:  $r = W\log \left( {1 + \frac{{|h{|^2}p}}{{W{\sigma ^2}}}} \right)$\cite{Xu2014}. A finite time horizon $T = 40$s is chosen, and the ``guard band'' \cite{Uysal-Biyikoglu2002} is set as 2s. We assume a Poisson arrival rate of the FIFO packets $\lambda = 2, 3$ packet/s, respectively. Further, we generate a string of packets, each of size $B_{\text{non}}$ which varies from $0.1$ KB to $2$ KB, with the arrival rate of 0.025 packet/s, and we take the first arrived packet in the string as the non-FIFO packet.
%The transmission time of FIFO and Non-FIFO packet are assumed to be $\lambda + 1$ and $\frac{\lambda + 1}{3}$, respectively.
The deadline for each FIFO packet (except the last one) is 4 second, and the non-FIFO packet must be transmitted within $t_{\text{non}}$ seconds of its arrival instant, where $t_{\text{non}}$ is one half of the difference between the deadline of its previous FIFO packet minus the arrival instant of the non-FIFO packet.

\begin{figure}[htb]
 %Requires
\centering
\includegraphics[width=.5\textwidth]{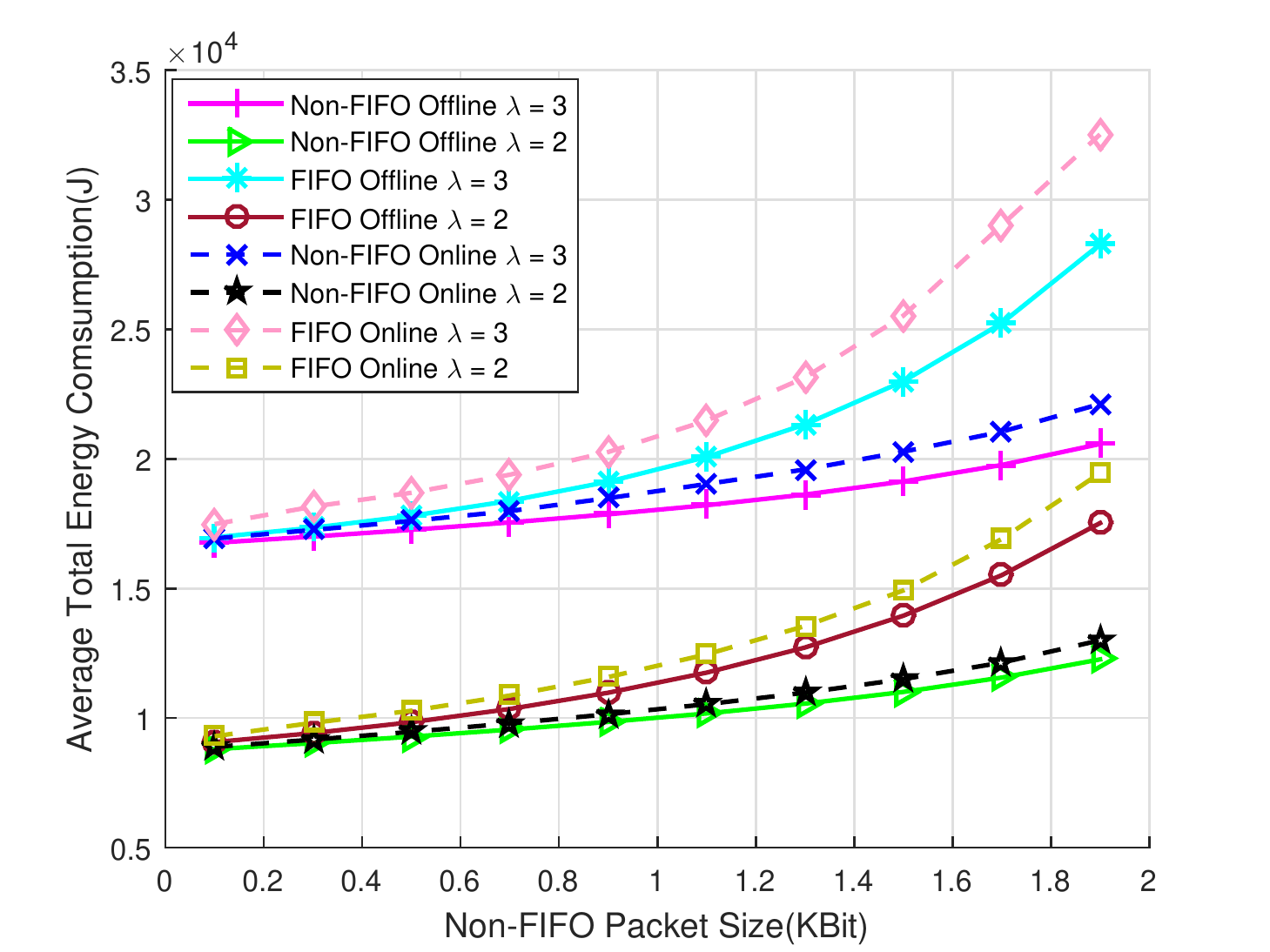}
\caption{Comparison of energy consumption for Non-FIFO offline (online) Scheduler and FIFO offline (online) Scheduler at different FIFO packet's arrival rate.}
\label{Simulation}
\end{figure}

Fig. \ref{Simulation} demonstrates the average total energy consumption difference between Non-FIFO offline (online) scheduler and FIFO offline (online) scheduler under different FIFO packet arrival rate $\lambda$. The results were averaged over 1000 independent simulations. As can be seen, No matter $\lambda = 2$ packet/s or $\lambda = 3$ packet/s, the optimal Non-FIFO offline algorithm proposed in this paper always outperforms the FIFO offline scheduler \cite{Zafer2009}.
%, yielding the lower bound of energy consumption for completing the transmission of all the packets.
With the increasing of the size of the Non-FIFO packet, the energy saving effect of the proposed algorithm is more and more obvious. Especially, when  $B_{\text{non}}=2$ KB, the proposed algorithm can save up to almost $40$ percent energy over the existing FIFO offline schedule \cite{Zafer2009}. In addition, when $\lambda$ increases from 2 packet/s to 3 packet/s, the average total energy consumption of all algorithms increases due to the fact that the data packets need to be transmitted at a higher rate during the finite time horizon. Lastly, it is also observed that the energy consumption with the proposed online algorithm is closed to that of the optimal offline scheme even without the knowledge of future arrivals, especially for the Non-FIFO online algorithm. Meanwhile, the Non-FIFO online algorithm also always outperforms the FIFO online scheme \cite{Zafer2009} in terms of energy saving, and the saving is more than 45 percent when $B_{\text{non}}=2$ KB.

\section{Conclusion}
We considered the problem of minimizing the transmission energy consumption with a non-FIFO packet over a point to point AWGN invariant channel under the feasibility constraints. We proposed a novel packet split and reorder process to transform the original problem into the problem of finding the optimal split factor of the packet that arrived before the non-FIFO packet. Assuming all arrival information is available, we formulated the problem using a calculus approach and modeled the feasibility constraints of data flow as cumulative curves. We proposed a strategy for finding the optimal departure curve which consisted of checking 4  possibilities by applying the ``String Tautening'' algorithm for FIFO packets. Then, the appealing and intuitive graphical visualization of the proposed transmission strategy was also given. By exploiting the optimality properties of the optimal departure curve in presence of a non-FIFO packet, we proved the optimality of our proposed strategy. Inspired by the optimal offline scheme, a practical energy efficient online scheduler is also studied which has a comparable performance to the proposed offline scheme. As part of our future work, we will study the same problem where there are \emph{multiple} Non-FIFO packets.

\section{Acknowledgement}
This work is partially supported by the National Basic Research Program of China(973 Program 2012CB316004), the National Natural Science Foundation of China
under Grants $61201170$, $61271208$ and $61221002$ and Qing Lan Project.

\appendices
\section{Proof of Lemma 1}\label{app_split_optimal}
Since all the packets $P_1, P_2, \cdots, P_{j-2}$ has deadline earlier than $t_{d,j}$ and arrive before $P_{j-1}$, according to previous results where there are no non-FIFO packet, without loss of generality, we may consider strategies where they should be transmitted before the transmission of packets $P_{j-1}$ and $P_j$ in its entirety. Similarly, packets $P_{j+1},\cdots, P_N$ arrive later than $t_{a,j}$, and has a deadline after the deadline of $P_{j-1}$, therefore, according to previous results when there are no non-FIFO packet, without loss of generality, we may consider strategies where they should be transmitted in their entirety after the transmission of $P_j$ and $P_{j-1}$.

The transmission of $P_j$ should be in one piece, as there is no incident during that causes the need to interrupt its transmission. So without loss of generality, we may assume that for any strategy, before the transmission of $P_j$, there may have been a part of $P_{j-1}$ that has been transmitted, and this is exactly the definition of the split factor $S_{j-1}$. After the transmission of $P_j$, the amount of $P_{j-1}$ transmitted is the remaining part. Thus, the optimal strategy can have the simple form of the split and reorder process. \hfill $\Box$

\section{Proof of Lemma 2 and 3}\label{app_S=0/B}
I. The proof of Lemma 2.

\begin{figure*}[htpb]
\centering
\includegraphics[width=1.0\textwidth]{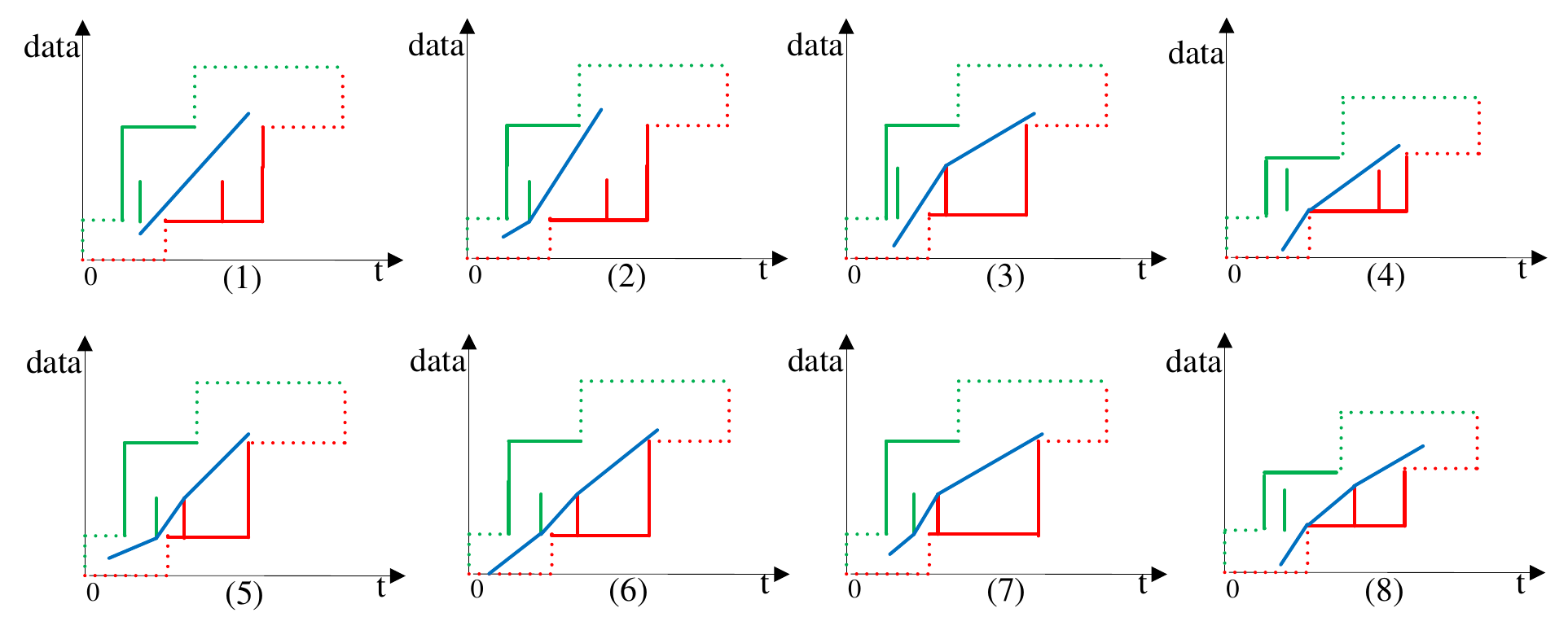}
\caption{All possible optimal departure curve scenarios of ${D^{SAR}_{opt}}(t;S_{j-1})$ when $S_{j-1} = 0$ according to the intersection number among ${D^{SAR}_{opt}}(t;0)$, $A^{SAR}\left( t_{a,j};{0} \right)$, $D^{SAR}_{\min }\left( t_{d,j};0 \right)$ and $D^{SAR}_{\min }\left( t_{d,j-2};0 \right)$.}
\label{AllCase_S=0}
\end{figure*}

We firstly prove the sufficient part. i.e., if ${D^{SAR}_{opt}}'(t_{j-2};0)\ge{D^{SAR}_{opt}}'(t_{j-1};0)$, then, $S_{j-1}^{opt} = 0$.

According to the number of intersection point among ${D^{SAR}_{opt}}(t;0)$, $A^{SAR}\left( t_{a,j};{0} \right)$, $D^{SAR}_{\min }\left( t_{d,j};0 \right)$ and $D^{SAR}_{\min }\left( t_{d,j-2};0 \right)$,  the optimal departure curve can only take one of the eight scenarios depicted in Figure \ref{AllCase_S=0}. In scenarios (1)(3)(4)(6)(7)(8), the optimal departure curves satisfy ${D^{SAR}_{opt}}'(t_{j-2};0)\ge{D^{SAR}_{opt}}'(t_{j-1};0)$, while in (2) and (5), ${D^{SAR}_{opt}}'(t_{j-2};0) <{D^{SAR}_{opt}}'(t_{j-1};0)$.

Similar to the proof of scenarios (1) in appendix \ref{app_0<S<B}, the optimal departure curve in scenarios (1) and (4) are the optimal policy since the optimal departure curve neither intersect $A^{SAR}\left( t_{a,j};{0} \right)$ nor $D^{SAR}_{\min }\left( t_{d,j};0 \right)$, and thus we can relax $P_j$'s causality and deadline constraints to that of $P_{j-1}$. Hence, the packet sequence turn into FIFO sequence and the optimal departure curve obtained by algorithm 1 is the optimal policy.

Similar to the proof of scenarios (5) in appendix \ref{app_0<S<B}, in scenarios (3)(6)(7)(8), when the value of split factor is increasing from $0$, it is obvious that the transmit-rate of $P_{j-1}^{sub1}$ is increasing from ${D^{SAR}_{opt}}^\prime \left( {{t_{j-2}};{0}} \right)$ while the transmit-rate of $P_{j-1}^{sub2}$ is decreasing from ${D^{SAR}_{opt}}^\prime \left( {{t_{j-1}};{0}} \right)$, i.e., ${D^{SAR}_{opt}}^\prime \left( {{t_{sub1}};{{S}_{j - 1}}} \right) \neq {D^{SAR}_{opt}}^\prime \left( {{t_{sub2}};{{S}_{j - 1}}} \right)$ hold when $S_{j-1} \in (0,B_{j-1})$. According to the proof of the necessary part in appendix \ref{app_0<S<B}, we can conclude that the optimal split factor does not lie in the interval $(0,B_{j-1})$. When the split factor's value is $B_{j-1}$, there are two possibilities to consider since $t_{a,j}<t_{a,j+1}$:
\\Case 1-$t_{d,j}<t_{a,j+1}$: There are must exist a ``idle'' period $[t_{d,j},t_{a,j+1}]$ without any transmission since $P_{j}$ must be transmitted before $t_{d,j}$ and the departure curve is non-decreasing function with time $t$. According to \cite{Uysal-Biyikoglu2002}, ``Non-Idling'' transmission is the necessary condition for optimal transmission. Hence, in this case, $S_{j-1}^{opt} \neq B_{j-1}$.
\\Case 2-$t_{d,j} \ge t_{a,j+1}$: In this case, it is obvious that the transmit-rate of $P_{j+1}$ is smaller than that of $P_{j-1}$, i.e., ${D^{SAR}_{opt}}'(t_{j+1};B_{j-1})<{D^{SAR}_{opt}}'(t_{j-1};B_{j-1})$. However, ${D^{SAR}_{opt}}'(t_{j+1};B_{j-1})<{D^{SAR}_{opt}}'(t_{j-1};B_{j-1})$ is not the sufficient condition for $S_{j-1}^{opt} = B_{j-1}$ and the proof is similar to the proof of necessary part for $S^{opt}= 0$, i.e., ${D^{SAR}_{opt}}'(t_{j-2};0)<{D^{SAR}_{opt}}'(t_{j-1};0)$ is not the sufficient condition for $S^{opt}= 0$, in the following part. Henceforth, the optimal departure curve in scenarios (3)(6)(7)(8) where ${D^{SAR}_{opt}}'(t_{j-2};0)\ge{D^{SAR}_{opt}}'(t_{j-1};0)$, are the optimal policy, i.e., $S_{j-1}^{opt} = 0$.

Next, we prove the necessary part. i.e., if $S^{opt}=0$, then ${D^{SAR}_{opt}}'(t_{j-2};0)\ge{D^{SAR}_{opt}}'(t_{j-1};0)$ or if ${D^{SAR}_{opt}}'(t_{j-2};0)<{D^{SAR}_{opt}}'(t_{j-1};0)$, then $S^{opt}\neq 0$.

We assume ${D^{SAR}_{opt}}'(t_{j-2};0)<{D^{SAR}_{opt}}'(t_{j-1};0)$, which corresponds to scenarios (2) and (5). According to \cite[Lemma 4]{Zafer2009}, $D^{SAR}_{opt}(t;0)$ must has a intersection with $A^{SAR}(t_{a,j};0)$. Since the packet sequence order of $\mathcal{P}^{SAR}(0)$ is $P_{1},\cdots,P_{j-2},P_{j},P_{j-1},\cdots,P_{N}$, we thus can denote the actual transmission duration of $P_{j-2}$, $P_{j}$ and $P_{j-1}$ as $(a_{j-2},t_{a,j}]$, $(t_{a,j},a_{j-1}]$ and $(a_{j-1},b_{j-1}]$, respectively. Meanwhile, we denote $\tau_1 = t_{a,j}- a_{j-2}$ and $\tau_2 = b_{j-1}- a_{j-1}$, respectively.

There must exist a split factor $S_{j-1}=\triangle>0$  that we can always construct a feasible departure curve $\hat{D}(t;S_{j-1})$, $t \in [0,T]$
\begin{equation}\label{}
  {\hat{D}'}(t;S_{j-1}) = \left\{ {\begin{array}{*{20}{c}}
                {{D^{SAR}_{opt}}'(t;0),} &{t\notin(a_{j-2},t_{a,j}]\cup(a_{j-1},b_{j-1}];}\\
                {\frac{{B_{j-2}+\triangle}}{\tau_1},} &{t\in(a_{j-2},t_{a,j}];}\\
                {\frac{{B_{j-1}-\triangle}}{\tau_2},} &{t\in(a_{j-1},b_{j-1}].}
                \end{array}} \right.
\end{equation}
such that $\frac{{B_{j-2}+\triangle}}{\tau_1}<\frac{{B_{j-1}-\triangle}}{\tau_2}$, and thus, $\triangle<\frac{{\tau_1}B_{j-1}-{\tau_2}B_{j-2}}{{\tau_1}+{\tau_2}}$.
Meanwhile, $P_{j-1}$ is thus split into two sub-packets $P_{j-1}^{sub1}$ and $P_{j-1}^{sub2}$ with size of $\triangle$ and $B_{j-1}-\triangle$, respectively. The packet sequence order of $\mathcal{P}^{SAR}(S_{j-1})$ is $P_{1},\cdots,P_{j-2},P_{j-1}^{sub1},P_{j-1},P_{j-1}^{sub2},\cdots,P_{N}$. In other words, this new constructed departure curve is the same as the optimal strategy in the assumption when $t\notin(a_{j-2},t_{a,j}]\cup(a_{j-1},b_{j-1}]$; the $j-2$-th packet and the first sub-packet of $j-1$-th packet are transmitted  in the duration $(a_{j-2},t_{a,j}]$ with the same rate $\frac{{B_{j-2}+\triangle}}{\tau_1}$, and the second sub-packet of $j-1$-th packet is transmitted in the duration $t\in(a_{j-1},b_{j-1}]$, i.e., the rate is $\frac{{B_{j-1}-\triangle}}{\tau_2}$. Since ${D^{SAR}_{opt}}'(t_{j-1};0)=\frac{B_{j-1}}{\tau_2}$ and ${D_{opt}}'(t_{j-2};0)=\frac{B_{j-2}}{\tau_1}$ , then, the following inequality established:
\begin{equation}\label{rateinequal}
  {D_{opt}}'(t_{j-2};0)<\frac{{B_{j-2}+\triangle}}{\tau_1}<\frac{{B_{j-1}-\triangle}}{\tau_2}<{D_{opt}}'(t_{j-1};0)
\end{equation}

\begin{figure}[htpb]
\centering
\includegraphics[width=.45\textwidth]{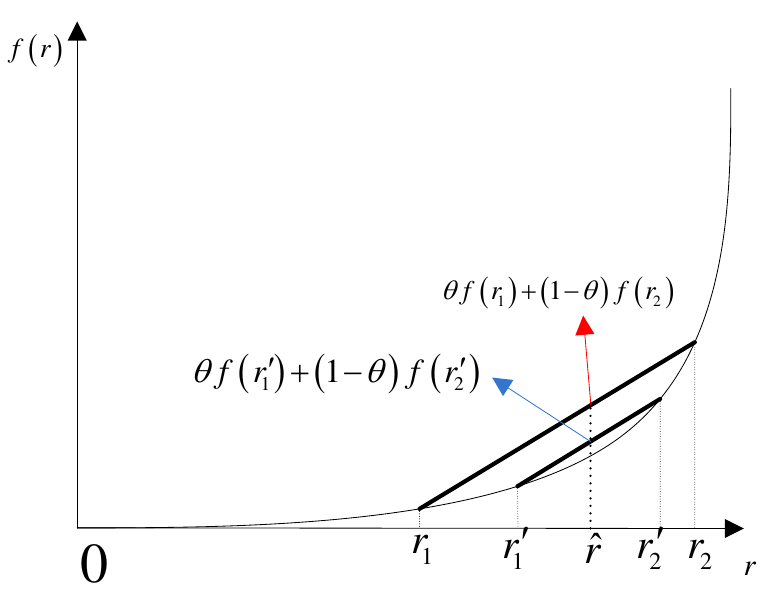}
\caption{$f(r)$ is increasing and convex in $r$.}
\label{Lemma1_f(r)}
\end{figure}

The energy consumption difference between $\hat{D}(t;S_{j-1})$, $t \in [0,T]$ and $D^{SAR}_{opt}(t;0)$, $t \in [0,T]$ is
\begin{equation}\label{energy_difference_S=0}
\begin{array}{l}
 E\left( {\hat D\left( t ;S_{j-1}\right)} \right) - E\left( {{D^{SAR}_{opt}}\left( {t;0} \right)} \right) \\
  = \int_{{a_{j - 2}}}^{{t_{a,j}}} {\left[ {f\left( {\hat D'\left( t ;S_{j-1}\right)} \right) - f\left( {{D^{SAR}_{opt}}^\prime \left( {t;0} \right)} \right)} \right]} dt \\
  + \int_{{a_{j - 1}}}^{{b_{j - 1}}} {\left[ {f\left( {\hat D'\left( t;S_{j-1} \right)} \right) - f\left( {{D^{SAR}_{opt}}^\prime \left( {t;0} \right)} \right)} \right]} dt \\
  = {\tau _1}\left[ {f\left( {\frac{{{B_{j - 2}} + \Delta }}{{{\tau _1}}}} \right) - f\left( {\frac{{{B_{j - 2}}}}{{{\tau _1}}}} \right)} \right] \\
  + {\tau _2}\left[ {f\left( {\frac{{{B_{j - 1}} - \Delta }}{{{\tau _2}}}} \right) - f\left( {\frac{{{B_{j - 1}}}}{{{\tau _2}}}} \right)} \right] \\
 \end{array}
\end{equation}
We let $\theta = \frac {\tau _1} {\tau _1 + \tau _2}$, $r_1 = {D^{SAR}_{opt}}'(t_{j-2};0)$, $r_2 = {D^{SAR}_{opt}}'(t_{j-1};0)$, $r'_1 = \hat{D}'(t;S_{j-1})$, where $t \in (a_{j-2},t_{a,j}]$; $r'_2 = \hat{D}'(t;S_{j-1})$, where $t \in (a_{j-1},b_{j-1}]$. Then, equation (\ref{energy_difference_S=0}) can be rewritten as:
\begin{equation}\label{energy_difference_S=0_equate}
\begin{array}{l}
 E\left( {\hat D\left( t ;S_{j-1}\right)} \right) - E\left( {{D^{SAR}_{opt}}\left( {t;0} \right)} \right) \\
  = {\tau _1}\left[ {f\left( {\frac{{{B_{j - 2}} + \Delta }}{{{\tau _1}}}} \right) - f\left( {\frac{{{B_{j - 2}}}}{{{\tau _1}}}} \right)} \right] \\
  + {\tau _2}\left[ {f\left( {\frac{{{B_{j - 1}} - \Delta }}{{{\tau _2}}}} \right) - f\left( {\frac{{{B_{j - 1}}}}{{{\tau _2}}}} \right)} \right] \\
  = (\tau_1+\tau_2)[\theta f(r'_1)+(1-\theta)f(r'_2)-\theta f(r_1)-(1-\theta)f(r_2)]
\end{array}
\end{equation}
Since $r_1 \tau_1 + r_2 \tau_2 = r'_1 \tau_1 + r'_2 \tau_2 $, then $\theta r_1 +(1-\theta)r_2 = \theta r'_1 +(1-\theta)r'_2 = \hat{r} \in (r'_1,r'_2)$. Combined the fact that $f(\cdot)$ is strictly increasing and convex function with the Inequality (\ref{rateinequal}), which are shown in Fig. \ref{Lemma1_f(r)}. According to  \cite{Boyd2004}, we have
\begin{equation}\label{energy_inequal}
  \theta f(r'_1)+(1-\theta)f(r'_2)< \theta f(r_1)-(1-\theta)f(r_2)
\end{equation}
(\ref{energy_inequal}) indicates that $E\left( {\hat D\left( t ;S_{j-1}\right)} \right) - E\left( {{D^{SAR}_{opt}}\left( {t;0} \right)} \right) <0$ and thus ${D^{SAR}_{opt}}'(t_{j-1};0) > {D^{SAR}_{opt}}'(t_{j-2};0)$ concludes that $S^{opt} \neq 0$. In turn, if $S^{opt} = 0$, then ${D^{SAR}_{opt}}'(t_{j-1};0) \le {D^{SAR}_{opt}}'(t_{j-2};0)$. The proof is completed.

II. The proof of Lemma 3.

Since the proof of necessary and sufficient condition in Lemma 3 is similar with that in Lemma 2, hence due to the limited space, the details in Lemma 3 are won't be covered in this paper. \hfill $\Box$

\section{Proof of Lemma 4}\label{app_0<S<B}
Firstly, we prove the necessary part.

Fig.3 show all the possible optimal departure curves when $S_{j-1} \in (0,B_{j-1})$, where ${D^{SAR}_{opt}}^\prime \left( {{t_{sub1}};{{S}_{j - 1}}} \right) = {D^{SAR}_{opt}}^\prime \left( {{t_{sub2}};{{S}_{j - 1}}} \right)$ in scenarios (1) and (5), and  ${D^{SAR}_{opt}}^\prime \left( {{t_{sub1}};{{S}_{j - 1}}} \right) \neq {D^{SAR}_{opt}}^\prime \left( {{t_{sub2}};{{S}_{j - 1}}} \right)$ in the remaining scenarios. Connect the two discrete time interval $\left( {{a^{sub1}_{j-1}},{b^{sub1}_{j-1}}} \right]$ and $\left( {{a^{sub2}_{j-1}},{b^{sub2}_{j-1}}} \right]$ which are respectively the transmission time interval for sub-packet $P_{j-1}^{sub1}$ and $P_{j-1}^{sub2}$, together, and shift the connected interval's starting point to 0. Meanwhile, we denote $\tau_{j-1} = (b^{sub1}_{j-1}-a^{sub1}_{j-1})+(b^{sub2}_{j-1}-a^{sub2}_{j-1})$. If the two sub-packets are transmitted at the same rate, i.e.,$\frac{B_{j-1}}{\tau_{j-1}}$, then, the energy consumption of these two packets are $\tau_{j-1}f(\frac{B_{j-1}}{\tau_{j-1}})$.

According to \cite{Gradshteyn2000}, Jensen's inequality can be expressed as:
 \begin{equation}\label{Jensen}
  \phi \left( {\frac{{\int_a^b {v\left( x \right)q\left( x \right)dx} }}{{\int_a^b {q\left( x \right)dx} }}} \right) \le \frac{{\int_a^b {\phi \left( {v\left( x \right)} \right)q\left( x \right)dx} }}{{\int_a^b {q\left( x \right)dx} }}
\end{equation}
We let $\phi \left(  \cdot  \right) = f\left(  \cdot  \right)$, $v\left(  \cdot  \right) = D'\left(  \cdot  \right)$, $q\left(  \cdot  \right) = 1$, $a=0$, $b=\tau_{j-1}$ and $x = t$, then we get:
\begin{equation}\label{EnergyDifference}
f\left( {\frac{{\int_0^{{\tau _{j-1}}} {\frac{{{B_{j-1}}}}{{{\tau _{j-1}}}}dt} }}{{\int_0^{{\tau _{j-1}}} {1dt} }}} \right) \le \frac{{\int_0^{{\tau _{j-1}}} {f\left( {{D'}\left( t \right)} \right)dt} }}{{\int_0^{{\tau _{j-1}}} {1dt} }}
\end{equation}
Through the equivalent transformation of (\ref{EnergyDifference}), we can get
\begin{equation}\label{}
{\tau _{j-1}}f\left( {\frac{{{B_{j-1}}}}{{{\tau_{j-1}}}}} \right) \le \int_0^{{\tau_{j-1}}} {f\left( {{D'}\left( t \right)} \right)dt}
\end{equation}
Therefore, in the optimal $S_{j-1}^{opt} \in (0,B_{j-1})$, the two sub-packets should be transmitted at a constant rate, i.e., ${D^{SAR}_{opt}}^\prime \left( {{t_{sub1}};{{S}^{opt}_{j - 1}}} \right) = {D^{SAR}_{opt}}^\prime \left( {{t_{sub2}};{{S}^{opt}_{j - 1}}} \right)$.

Next, we will prove the sufficient part, i.e., if $S_{j-1} \in (0,B_{j-1})$, where ${D^{SAR}_{opt}}^\prime \left( {{t_{sub1}};{{S}_{j - 1}}} \right) = {D^{SAR}_{opt}}^\prime \left( {{t_{sub2}};{{S}_{j - 1}}} \right)$, then, $S_{j-1}$ is optimal. Hence, we only need to discuss scenarios (1) and (5).

The optimal departure curve in scenario (1) is obviously the optimal policy since $D^{SAR}_{opt}(t, S_{j-1})$ neither intersect $A^{SAR}\left( t_{a,j};{S_{j - 1}} \right)$ nor $D^{SAR}_{\min }\left( t_{d,j};S_{j - 1} \right)$, hence we can relax the causality and deadline constraints of $P_j$ to that of $P_{j-1}$. Henceforth, the sequence turn into a FIFO sequence and the optimal departure curve by algorithm 1 in Scenario (1) is the optimal policy.

As for scenario (5), when the value of split factor is decreasing from $S_{j-1}$, the transmit-rate of $P_{j-1}^{sub1}$ is decreasing while the transmit-rate of $P_{j-1}^{sub2}$ is increasing, i.e., ${D^{SAR}_{opt}}^\prime \left( {{t_{sub1}};{{S}_{j - 1}}} \right) \neq {D^{SAR}_{opt}}^\prime \left( {{t_{sub2}};{{S}_{j - 1}}} \right)$ hold when the split factor lies in the interval $(0,S_{j-1})$. According to the proof of the necessary part above in appendix \ref{app_0<S<B}, we can conclude that the optimal split factor does not lie in the interval $(0,S_{j-1})$. In addition, when the split factor's value is 0, there are two possibilities to consider since $t_{d,j}>t_{d,j-2}$:
\\Case 1 - $t_{a,j}>t_{d,j-2}$: There are must exist a ``idle'' period $[t_{d,j-2},t_{a,j}]$ without any transmission since $P_{j-2}$ must be transmitted before $t_{d,j-2}$ and the departure curve is non-decreasing function with time $t$. According to \cite{Uysal-Biyikoglu2002}, ``Non-Idling'' transmission is the necessary condition for optimal transmission. Hence, in this case, $S_{j-1}^{opt} \neq 0$.
\\Case 2 - $t_{a,j} \le t_{d,j-2}$: In this case, it is obvious that the transmit-rate of $P_{j-2}$ is smaller than that of $P_{j-1}$, i.e., ${D^{SAR}_{opt}}'(t_{j-2};0)<{D^{SAR}_{opt}}'(t_{j-1};0)$. However, ${D^{SAR}_{opt}}'(t_{j-2};0)<{D^{SAR}_{opt}}'(t_{j-1};0)$ is not the sufficient condition for $S_{j-1}^{opt} = 0$. The details of the proof are shown in the proof of necessary part for optimality in Lemma 2.

In addition, By similar analysis, we can learn that the value of split factor lies in $(S_{j-1},B_{j-1}]$ is not optimal.

Therefore, the optimal departure curve in scenario (5) where ${D^{SAR}_{opt}}^\prime \left( {{t_{sub1}};{{S}_{j - 1}}} \right) = {D^{SAR}_{opt}}^\prime \left( {{t_{sub2}};{{S}_{j - 1}}} \right)$, is the optimal policy and $S_{j-1}^{opt} = S_{j-1}$, where $S_{j-1} \in (0,B_{j-1})$.\hfill $\Box$

\bibliographystyle{IEEEtran}
\bibliography{Myreference}

\begin{thebibliography}{10}
\providecommand{\url}[1]{#1}
\csname url@samestyle\endcsname
\providecommand{\newblock}{\relax}
\providecommand{\bibinfo}[2]{#2}
\providecommand{\BIBentrySTDinterwordspacing}{\spaceskip=0pt\relax}
\providecommand{\BIBentryALTinterwordstretchfactor}{4}
\providecommand{\BIBentryALTinterwordspacing}{\spaceskip=\fontdimen2\font plus
\BIBentryALTinterwordstretchfactor\fontdimen3\font minus
  \fontdimen4\font\relax}
\providecommand{\BIBforeignlanguage}[2]{{%
\expandafter\ifx\csname l@#1\endcsname\relax
\typeout{** WARNING: IEEEtran.bst: No hyphenation pattern has been}%
\typeout{** loaded for the language `#1'. Using the pattern for}%
\typeout{** the default language instead.}%
\else
\language=\csname l@#1\endcsname
\fi
#2}}
\providecommand{\BIBdecl}{\relax}
\BIBdecl

\bibitem{Berry2002}
R.~Berry and R.~Gallager, ``Communication over fading channels with delay
  constraints,'' \emph{IEEE Transactions on Information Theory}, vol.~48,
  no.~5, pp. 1135--1149, 2002.

\bibitem{Uysal-Biyikoglu2002}
E.~Uysal-Biyikoglu, B.~Prabhakar, and A.~El~Gamal, ``Energy-efficient packet
  transmission over a wireless link,'' \emph{IEEE/ACM Transactions on
  Networking}, vol.~10, no.~4, pp. 487--499, 2002.

\bibitem{Zafer2009}
M.~Zafer and E.~Modiano, ``A calculus approach to energy-efficient data
  transmission with quality-of-service constraints,'' \emph{IEEE/ACM
  Transactions on Networking}, vol.~17, no.~3, pp. 898--911, 2009.

\bibitem{Chen2008}
W.~Chen, M.~Neely, and U.~Mitra, ``Energy-efficient transmissions with
  individual packet delay constraints,'' \emph{IEEE Transactions on Information
  Theory}, vol.~54, no.~5, pp. 2090--2109, 2008.

\bibitem{Nan2014a}
Z.~Nan, X.~Wang, and W.~Ni, ``Energy-efficient transmission of delay-limited
  bursty data packets under non-ideal circuit power consumption,'' in
  \emph{Communications (ICC), 2014 IEEE International Conference on}, 2014, pp.
  4957--4962.

\bibitem{Jin2014}
Y.~Jin, J.~Xu, and L.~Qiu, ``Energy-efficient scheduling with individual packet
  delay constraints and non-ideal circuit power,'' \emph{Journal of
  Communications and Networks}, vol.~16, no.~1, pp. 36--44, 2014.

\bibitem{Xu2014}
J.~Xu and R.~Zhang, ``Throughput optimal policies for energy harvesting
  wireless transmitters with non-ideal circuit power,'' \emph{Selected Areas in
  Communications, IEEE Journal on}, vol.~32, no.~2, pp. 322--332, 2014.

\bibitem{Bai2011}
Q.~Bai, J.~Li, J.~Nossek \emph{et~al.}, ``Throughput maximizing transmission
  strategy of energy harvesting nodes,'' in \emph{Cross Layer Design (IWCLD),
  2011 Third International Workshop on}.\hskip 1em plus 0.5em minus 0.4em\relax
  IEEE, 2011, pp. 1--5.

\bibitem{Ozel2011}
O.~Ozel, K.~Tutuncuoglu, J.~Yang, S.~Ulukus, and A.~Yener, ``Transmission with
  energy harvesting nodes in fading wireless channels: Optimal policies,''
  \emph{Selected Areas in Communications, IEEE Journal on}, vol.~29, no.~8, pp.
  1732--1743, 2011.

\bibitem{Yang2012}
J.~Yang and S.~Ulukus, ``Optimal packet scheduling in an energy harvesting
  communication system,'' \emph{Communications, IEEE Transactions on}, vol.~60,
  no.~1, pp. 220--230, 2012.

\bibitem{Boyd2004}
S.~Boyd and L.~Vandenberghe, \emph{Convex optimization}.\hskip 1em plus 0.5em
  minus 0.4em\relax Cambridge university press, 2004.

\bibitem{Gradshteyn2000}
I.~Gradshteyn and I.~Ryzhik, ``Table of integrals, series, and products 6th edn
  (san diego, ca: Academic),'' 2000.

\end{thebibliography}

\end{document}